# Student Preferences for Online Interaction Platforms in Blended Learning: A Mixed-Methods Study

Lois Fajuyigbe[1], Kaisu Mumuni[1] and Felix Nti Koranteng[1,2,*]

[1] Accra Institute of Technology, Accra, Ghana
[2] Eindhoven University of Technology, Eindhoven, The Netherlands

**Abstract**

As higher education increasingly adopts blended learning, understanding students' preferences for online interaction platforms becomes critical for effective course delivery and engagement. This study investigates the platforms undergraduate students prefer for academic communication and explores the underlying reasons for these choices. Data were collected from 37 students enrolled in two summer courses at a Ghanaian university using a structured questionnaire consisting of both closed and open-ended items. Quantitative results revealed a strong preference for instant messaging platforms such as WhatsApp and Telegram over institutional learning management systems. Qualitative content analysis of the open-ended responses identified five key factors influencing platform preference: convenience and familiarity, ease of use, accessibility, popularity among peers, and support for real-time interactions. These findings highlight a significant mismatch between students' communication habits and institutional platform offerings. The study highlights the importance of aligning digital learning strategies with students' lived digital experiences to enhance interaction, collaboration, and learner satisfaction in blended learning environments.

## 1. Introduction

In recent years, the integration of digital technology into higher education has become not just a supplementary tool but a defining feature of modern instructional delivery. As universities worldwide adapt to new modes of teaching and learning, blended learning, an approach that combines traditional classroom instruction with digital engagement, has emerged as a widely adopted model [1]. This shift has brought both opportunities and challenges for students and educators, particularly in how learners interact and collaborate within these hybrid environments.

Central to the success of blended learning is the extent to which students actively participate in online interactions. Digital tools are intended to extend engagement beyond the classroom and offer channels for communication, feedback, and peer collaboration [2]. However, despite the availability of institutional learning platforms such as learning management systems, many educators continue to face difficulties in fostering meaningful student interaction online [3]. This raises critical questions about whether the tools provided by institutions are aligned with the communication preferences and habits of today's learners.

---

In most blended learning contexts, interaction is no longer limited to official course platforms. Students increasingly rely on alternative, often informal, digital communication tools (such as WhatsApp, Telegram, or Discord) for discussing assignments, collaborating on projects, and seeking clarification [4]. These preferences suggest a possible disconnect between institutional expectations and student behavior, which could have implications for course design, instructional strategies, and the overall effectiveness of blended learning environments. Understanding the factors that shape students' choices of digital platforms for academic interaction is, therefore, an important area of inquiry.

This study seeks to address this gap by exploring university students' preferences for online interaction platforms within the context of blended learning. It aims to identify the tools students are most inclined to use for academic collaboration and to understand the features or perceptions that drive these choices.

## 2. Background

The rapid advancement of digital technologies has significantly transformed various aspects of daily life, including education. One notable shift is the widespread adoption of online components in teaching and learning processes. Among these innovations, blended learning has emerged as a prominent instructional approach, combining face-to-face teaching with digital technologies [5]. The core objective of blended learning is to enhance communication and collaboration between students and instructors through supplementary digital platforms [2]. This mode of learning creates more dynamic environments where students and teachers can engage in meaningful discussions, exchange feedback, and share ideas effectively [6]. Research has shown that blended learning fosters greater interactivity and increases contact hours compared to traditional classroom settings [1], [7]. Moreover, the additional academic and peer support offered in blended settings has been found to boost student engagement and improve academic performance [8].

Building on this evidence, many universities have enhanced their traditional Learning Management Systems (LMS) by integrating emerging technologies such as discussion boards, digital whiteboards, and video conferencing tools to facilitate communication among students, peer groups, and instructors [3]. Despite these advancements, several studies report low levels of student interaction on these platforms [3], [9]. For example, Sarker et al. [9] found that course-related interactions among students remain minimal within LMS environments. Similarly, Budu [10] observed that students rarely respond to questions or content posted by instructors. Zhu and Dawson [11] further characterized students as "lurkers," noting that they often access course platforms solely to consume content rather than to engage in meaningful discussions. This limited interaction is largely attributed to students' perception of LMS platforms as repositories for course materials rather than as interactive communication tools [12]. Supporting this view, Ansong-Gyimah [13] reported that the majority of students primarily used the platform to download learning resources, with minimal engagement in collaborative or dialogic activities.

In relation to the above discussions, it is plausible that students may prefer alternative platforms for peer interaction over those officially provided by universities for course-related communication. This highlights the influence of user preferences on the adoption and effective use of technological tools in learning environments [4]. According to the Task-Technology Fit (TTF) theory [14], users are more likely to adopt a technology when they perceive it as well-suited to the task at hand. Conversely, technologies perceived as misaligned with a given task tend to be underutilized. For example, Koranteng et. al. [15] found that students do not use social networking sites such as Facebook for educational purposes because they perceive them as entertainment tools. However, existing literature has paid limited attention to students' specific platform preferences for academic interaction. Instead, much of the research has concentrated on other motivational factors such as individual motivation [16], self-efficacy [17], trust [18], and social influence [19], leaving a gap in understanding the role of platform preference in shaping online engagement.

To address the persistent challenge of low student engagement on institutional learning platforms, this study aims to explore university students' preferences for online interaction platforms within a blended learning context. Specifically, the objective of the current study is to identify which digital communication tools students gravitate toward for course-related interactions and to understand the factors that influence these preferences. Understanding these preferences can help educators and instructional designers select communication platforms that better align with students' habits and expectations, thereby improving participation, collaboration, and learning outcomes in blended environments.

## 3. Method

This study employed a qualitative research approach to explore university students' preferences for online interaction platforms in a blended learning context. Data were collected through an online survey designed to elicit students' preferred communication platforms and their underlying reasons for these choices. Responses were analyzed using content analysis to identify common themes and patterns related to platform preference.

### 3.1. Materials

Data were collected using a structured survey questionnaire developed and administered through Google Forms. The questionnaire consisted of both closed-ended and open-ended questions. Students were first asked to select their preferred online platform for course-related interaction from a list of commonly used platforms, including the institution's learning management systems (LEMASS), WhatsApp, Telegram, Google Classroom, etc. Following the selection, an open-ended question prompted students to explain the rationale behind their choice. Additional items were included to gather demographic information, including gender, age, and digital skills. The survey was designed to be user-friendly and accessible via mobile or desktop devices, allowing students to complete it at their convenience.

### 3.2. Procedure

The survey was administered at the end of the summer session, during the final week of instruction for both courses. Prior to distribution, the purpose of the study was communicated to students through class announcements on the institutional LMS and class WhatsApp groups. Students were informed that participation was voluntary, responses would remain anonymous, and data collected would be used solely for academic research purposes.

The instructor posted a link to the Google Forms questionnaire on both communication platforms. Students were encouraged to complete the survey at a time and location convenient to them, using either a smartphone or computer. The survey consisted of approximately 8 items and took an average of 5–7 minutes to complete. Students first responded to demographic questions, including their gender, age, and self-assessed digital skills. Next, they selected their preferred platform for course interaction from a predefined list. An open-ended text box then prompted them to explain why they chose that particular platform. At the end of the survey, participants were thanked for their participation and debriefed. No incentives or remuneration were offered.

### 3.3. Participants

Participants in this study were undergraduate students enrolled in two blended learning courses (i.e., IT Research Methods and Introduction to Artificial Intelligence) offered during a summer session at a university in Ghana. A convenience sampling technique was used to recruit students, based on their availability and enrollment in the selected courses. Out of a total of 55 students registered across both courses, 37 students completed the survey, resulting in a response rate of approximately 67%. The sample consisted of 19 males (51.4%) and 18 females (48.6%). Participants' ages ranged from 18 to 27 years, with a mean age of 21.9 years and a standard deviation of 2.4 years. In addition to demographic data, participants were asked to self-assess their digital skills. The majority (81%) rated their digital proficiency as either good or very good, indicating a generally high level of comfort with the use of digital tools for communication and learning. Participation in the study was voluntary, and all responses were anonymous to protect the confidentiality of the participants.

## 4. Data Analysis

A qualitative content analysis approach was employed to systematically examine and interpret students' responses to both closed-ended and open-ended survey items. The aim was to uncover recurring patterns and underlying factors that influenced students' preferences for online interaction platforms within a blended learning context. For the closed-ended items, responses were first tabulated to identify the frequency of selection for each listed platform (e.g., LEMASS, WhatsApp, Telegram, Google Classroom, etc.). This provided a quantitative overview of students' general platform preferences, which served as the basis for further qualitative interpretation.

The open-ended responses were analyzed using an inductive coding process. To enhance the reliability of the findings, two researchers independently coded the responses. These codes represented specific reasons or justifications given by students for their platform choices (such as "easy to use," "always on my phone," "everyone is on it," or "better for group work"). Similar codes were then grouped into broader, emergent themes. The researchers' initial coding results were then compared and discussed to resolve discrepancies and reach consensus on thematic labels. After reaching an agreement on the coded data, the researchers synthesized the results into broader categories.

Through this iterative review, five dominant themes emerged: *convenience and familiarity*, *ease of use*, *accessibility*, *Popularity,* and *Real-Time Interactions*. For example, comments about the platform being simple to navigate or convenient were classified under "ease of use". The final thematic structure enabled the researchers to interpret not only which platforms students preferred, but more importantly, why those preferences existed. These thematic insights inform the interpretation and discussion presented in the following section.

## 5. Data Analysis

### 5.1. Platform Preferences

The analysis of student responses revealed clear preferences for certain digital platforms used in course-related interactions. Out of the 37 students who participated in the survey, WhatsApp emerged as the most preferred platform, followed by Telegram, Google Classroom, LEMASS, and other platforms (see Figure 1). Specifically, a significant majority of students (62%) selected WhatsApp as their primary platform for academic interaction. Telegram was the second most selected platform (19%), followed by Google Classroom (11%). The Institutional platform (i.e., LEMASS), despite being officially promoted for academic engagement, was only selected by a small fraction of students (5%). A few respondents (3%) indicated a preference for other platforms such as Microsoft Teams, Zoom, and Google Meet. These findings suggest that students are more inclined to use informal, mobile-based messaging apps over institutional platforms for communication, collaboration, and academic support.

### 5.2. Thematic Reasons for Platform Choice

The qualitative analysis of students' open-ended responses revealed five key themes that explain why certain platforms were preferred for online interaction within blended learning environments: convenience and familiarity, ease of use, accessibility, popularity, and real-time interactions. These themes illustrate how students' platform choices were influenced by both technical features and social factors.

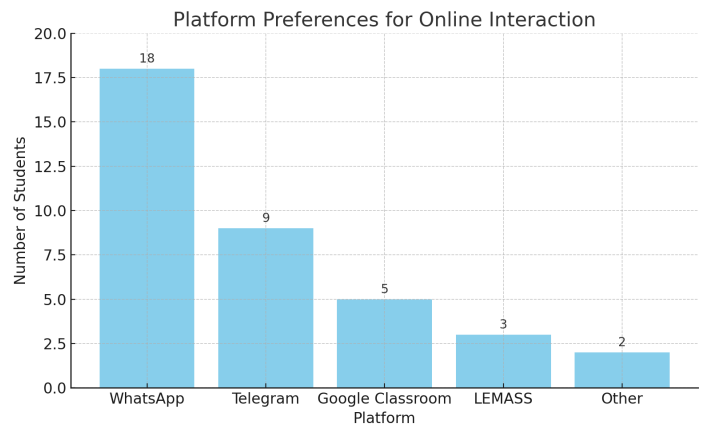

Figure 1. Platform Preferences for Online Interaction.

#### 5.2.1. Convenience and Familiarity

Students often selected platforms they were already using regularly in their daily lives, suggesting that habitual use influenced their comfort and confidence. WhatsApp and Telegram, in particular, were cited as familiar tools that did not require additional setup or learning curves. For instance, one student explained, "*I already use WhatsApp for most of my communications, so it's easier to keep track of class messages without having to check another app*" [PP21]. This familiarity made it more convenient to integrate academic discussions into their existing routines.

### 5.2.2. Ease of Use

Ease of navigation and straightforward interfaces were also critical factors. Students appreciated platforms that allowed them to quickly read messages, send responses, or share documents without technical complications. WhatsApp and Telegram were frequently described as "simple" and "user-friendly." *One respondent commented, "It's much easier to use than LEMASS. I don't get lost or miss messages*" [PP7]. This theme reflects how user experience design directly impacts platform preference in educational settings.

### 5.2.3. Accessibility

Several students cited practical limitations, such as limited data or unreliable internet access, as reasons for choosing more lightweight and mobile-friendly platforms. Institutional platforms like LEMASS or Google Classroom were perceived as data-intensive or requiring laptops, which some students found restrictive. One participant noted, "*I can check WhatsApp on my phone anytime, even with low data. LEMASS takes too long to load*" [PP34]. Accessibility, therefore, was not only about device compatibility but also about overcoming connectivity and economic barriers.

### 5.2.4. Popularity among Peers

The collective behavior of peers strongly shaped individual preferences. Students were more inclined to choose platforms where most of their classmates were active, allowing for broader participation and quicker responses. One student shared, "*Most of my classmates are on Telegram already. It's easier to follow the discussion when everyone is there*" [PP10]. This theme highlights the role of social dynamics in shaping the effectiveness of communication tools for academic collaboration.

### 5.2.5. Real-Time Interactions

Finally, the capacity for immediate feedback and spontaneous discussion emerged as a significant reason behind students' choices. Platforms like WhatsApp enabled instant messaging, voice notes, and even quick polls, allowing students to engage in real-time. One respondent explained, "*When I ask a question in WhatsApp, someone usually replies within minutes. That doesn't happen on LEMASS*" [PP11]. This immediacy enhanced students' sense of connection and responsiveness, which they found motivating and supportive in their learning process.

## 6. Discussion

The findings of this study offer important insights into students' preferences for online interaction platforms within a blended learning context. From the findings, WhatsApp and Telegram emerged as the most favored platforms, followed by Google Classroom, LEMASS, and other tools. The preference for informal communication platforms over institutional platforms highlights a significant shift in how students approach digital learning tools. The findings suggest that students' preferences are rooted in convenience, digital habits, and social interaction rather than formal structures.

These results can be understood through several theoretical lenses. For instance, the Media Richness Theory [20] suggests that people prefer communication channels that support immediate feedback, personalization, and rich information exchange. Platforms such as WhatsApp and Telegram support these characteristics. These platforms allow users to exchange text, voice notes, videos, and documents instantly, which may make them more attractive than formal LMS platforms like LEMASS that are often slower, less intuitive, and more rigid in design. In a similar vein, Zhang [21] posits that individuals actively choose media that fulfill their specific needs. For students in this study, these needs included quick access to peers and instructors, ease of navigation, and minimal data consumption. This helps explain why platforms such as WhatsApp and Telegram, which are used daily, were overwhelmingly preferred. Students viewed these platforms as a natural extension of their existing digital behaviors, eliminating the learning curve associated with newer or less familiar systems.

Accessibility also played a crucial role in shaping preferences. Several students noted that platforms like WhatsApp and Telegram worked well on low-end smartphones and consumed less data than institutional platforms. This reflects broader issues of digital equity in higher education, especially in contexts where internet connectivity, cost, and device ownership remain uneven. As Martin et. al. [22] point out, students' choices of learning tools are often conditioned by their socio-technical environment. When bandwidth and hardware limitations are a concern, lighter, mobile-optimized tools offer a more inclusive path to participation.

Another key factor was the influence of peers. Many respondents highlighted that their classmates were already active on certain platforms, making them more effective for coordination and feedback. This aligns with the Social Cognitive Theory [23], which emphasizes that people's behaviors and decisions are shaped by social interactions and observational learning. Platforms like WhatsApp and Telegram not only allowed for quick updates but also fostered a sense of community, peer support, and shared responsibility, all of which contribute to sustained engagement in learning activities. This is supported by prior studies such as Deng and Tavares [24] and Jacobs et. al. [25], who observed that social presence is a critical factor in learners' online engagement and satisfaction.

The preference for real-time communication further underscores the importance of immediacy in blended learning environments. Students valued the ability to send and receive messages instantly, clarify doubts, and share learning materials without delay. This real-time interaction is essential to building social presence. (Platforms that support synchronous or near-synchronous interaction enhance learners' emotional and cognitive investment, contributing to deeper learning outcomes.

These findings have significant implications for institutional policy and practice. The fact that students overwhelmingly favor non-institutional platforms suggests a disconnect between formal learning technologies and students' digital realities. While institutional LMSs offer structure, content control, and assessment integration, they may fall short in supporting the dynamic, informal interactions that many students rely on to stay engaged. As Almarzooq [26] argues, educational institutions may benefit from incorporating or integrating familiar communication platforms into their pedagogical strategies, thus bridging the gap between institutional design and student preferences. Rather than viewing informal platforms as distractions or risks, educators might consider how these tools can be harnessed to enrich the learning experience.

## 7. Conclusion and Future Work

This study examined the online platform preferences of undergraduate students engaged in blended learning and explored the underlying reasons for their choices. The findings reveal a strong preference for widely used messaging applications like WhatsApp and Telegram over institutional platforms, driven largely by factors such as convenience, familiarity, ease of use, accessibility, popularity among peers, and support for real-time interactions. These preferences underscore a growing reliance on informal digital tools that align more closely with students' everyday communication habits and technological realities.

The results highlight a critical gap between students' digital practices and the tools provided by educational institutions. To foster greater engagement and improve communication within blended learning environments, educators and administrators must consider integrating familiar, user-friendly platforms into course delivery while maintaining necessary academic structures and safeguards.

Future research could explore how such integrations influence learning outcomes and student satisfaction across different contexts. Additionally, institutional policies must evolve to support more flexible, inclusive digital learning environments that reflect the diverse needs and behaviors of today's learners.